\documentclass[draftcls,10pt]{ismrm}
\usepackage{endfloat}
\usepackage{makeidx}  % allows for indexgeneration
\usepackage{amsmath}
\usepackage{graphicx}
\usepackage{array}
\usepackage{bm}
\usepackage{color}
\usepackage{amssymb}
\usepackage{amsmath,graphicx,subfigure}
\usepackage{widetext}
\usepackage{hyperref}

\def\W{{\cal W}}

\def\F{{\cal F}}
\makeatletter
\renewcommand\tagform@[1]{\maketag@@@{\ignorespaces#1\unskip\@@italiccorr}}
\makeatother

\def\bk{$\boldsymbol k$}
\def\bx{$\boldsymbol x$}

\begin{document}

{\noindent \huge Computation of Exact g-Factor Maps in 3D GRAPPA Reconstructions (Submitted to Magnetic Resonance in Medicine}

\vspace{0.5cm}

% ----- AUTHORS -----

{ \noindent \large {I\~{n}aki Rabanillo-Viloria$^1$}, {Ante Zhu$^{2,3}$}, {Santiago Aja-Fern\'andez$^1$}, {Carlos Alberola-L\'opez$^1$}, and \\{Diego Hernando$^{3,4}$}

\vspace{0.25cm}

{\noindent \small
$^1$Laboratorio de Procesado de Imagen, Universidad de Valladolid, Valladolid, Spain.\\
$^2$Department of Biomedical Engineering, University of Wisconsin-Madison, Madison, WI.\\
$^3$Department of Radiology, University of Wisconsin-Madison, Madison, WI.\\
$^4$Department of Medical Physics, University of Wisconsin-Madison, Madison, WI.}

\vspace{0.5cm}

% ----- Corresponding author -----
{\bf \noindent \large Corresponding Author: }\\
I\~{n}aki Rabanillo-Viloria,\\
Laboratorio de Procesado de Imagen,\\
Universidad de Valladolid,\\
ETSI de Telecomunicaci\'on, Campus Miguel Delibes,\\
47011 Valladolid, Spain.\\
Email: irabvil@lpi.tel.uva.es\\
Phone: +34 983 423660, ext. 5590

\vspace{0.5cm}


\vspace{0.5cm}

{\noindent \bf Grant sponsor:} Spanish Ministerio de Ciencia e Innovación, {\bf Grant number:} TEC2013-44194P.\\
{\noindent \bf Grant sponsor:} Spanish Ministerio de Ciencia e Innovación, {\bf Grant number:} TEC 2014-57428.
{\noindent \bf Grant sponsor:} Spanish Ministerio de Ciencia e Innovación, {\bf Grant number:} TEC2017-82408-R.
{\noindent \bf Grant sponsor:} Junta de Castilla y Le\'on, {\bf Grant number:} VA069U16.\\
{\noindent \bf Grant sponsor:} Spanish Ministerio de Econom\'ia y Competitividad FPI Fellowship Program, {\bf Grant number:} BES-2014-069524.

\vspace{0.5cm}

{\bf \noindent \large Word count:} 5971

\clearpage

\begin{abstract}

{\bf \noindent Purpose:} To characterize the noise distributions in 3D-MRI accelerated acquisitions reconstructed with GRAPPA using an exact noise propagation analysis that operates directly in \bk--space. 

{\bf \noindent  Theory and Methods:} We exploit the extensive symmetries and separability in the reconstruction steps to account for the correlation between all the acquired \bk--space samples. Monte Carlo simulations and multi-repetition phantom experiments were conducted to test both the accuracy and feasibility of the proposed method; an \emph{in--vivo} experiment  was performed to assess the applicability of our method to clinical scenarios.

{\bf \noindent Results:} Our theoretical derivation shows that the direct \bk--space analysis renders an exact noise characterization under the assumptions of stationarity and uncorrelation in the original \bk--space. Simulations and phantom experiments provide empirical support to the theoretical proof. Finally, the \emph{in--vivo} experiment demonstrates the ability of the proposed method to assess the impact of the sub-sampling pattern on the  overall noise behavior.

{\bf \noindent Conclusions:} By operating directly in the \bk--space, the proposed method is able to provide an exact characterization of noise for any Cartesian pattern sub-sampled along the two phase--encoding directions. Exploitation of the symmetries and separability into independent blocks through the image reconstruction procedure allows us to overcome the computational challenges related to the very large size of the covariance matrices involved.
\end{abstract}

%%%%%%%%%%%%%%%%%%%%%%%%%%%%%%%%%%%%%%%%
%INTRODUCTION--------------
%%%%%%%%%%%%%%%%%%%%%%%%%%%%%%%%%%%%%%%

\section{Introduction}

Characterization of noise statistics in MR images is essential for multiple applications including quality assurance \cite{Krissian09,Aja13}, protocol optimization \cite{Thunberg07,Saritas11}, and tailoring of subsequent post--processing steps \cite{Rabanillo16,Ghugre05,Veraart11}. For fully--sampled acquisitions, the noise distribution is spatially homogeneous, and is well modelled by a zero--mean spatially uncorrelated independent and identically distributed (IID) complex Gaussian process for each coil \footnote{If the data are acquired by several receiving coils, the multi-coil noise can be characterized by its covariance matrix}. However, in the presence of parallel MRI (pMRI) acceleration, the noise distribution (e.g. the pixel-wise noise standard deviation) generally has a strong spatial dependence \cite{Pruessmann99,AjaMRM11}. For these reasons, there is a clinical and research need for accurate estimation of the pixel-wise maps of noise standard deviation (i.e. `noise maps'), that is widely applicable to common pMRI methods.

In particular, commonly used pMRI methods where the reconstruction takes place in the \bk--space, such as generalized autocalibrating partially parallel acquisitions (GRAPPA) ~\cite{Griswold02}, introduce correlations in the \bk--space that propagate into the \bx--space. For GRAPPA reconstructions, a direct noise propagation analysis requires operating with very large covariance matrices~\cite{Robson08} and renders this analysis challenging. In recent work, we have shown that, by exploiting the extensive symmetries and separability in each step of the reconstruction, a computationally efficient and exact noise analysis can be obtained for 2D  Cartesian acquisitions \cite{Rabanillo17}, where acceleration is applied along a single (phase-encoding) dimension. However, no feasible and exact noise analysis has been reported for 3D Cartesian acquisitions, which are commonly obtained with acceleration in two dimensions.

Indeed, 3D acquisitions are commonplace in clinical and research MRI exams, both in brain and body imaging applications. Compared to 2D imaging, 3D acquisitions enable high (eg: isotropic) spatial resolution with high signal-to-noise ratio (SNR) efficiency. Further, pMRI techniques can be combined with 3D imaging by sub-sampling the \bk--space along both phase--encoding directions \cite{Blaimer06}, resulting in highly flexible sub-sampling schemes. However, GRAPPA reconstruction in 3D imaging introduces correlations in \bk--space across the three dimensions, notably increasing the size of the matrices involved in the noise propagation analysis. For this reason, noise characterization in 3D imaging is substantially more challenging than in 2D imaging. 

Two previously proposed approaches to estimate the noise maps are given by the Monte--Carlo ~\cite{Robson08} and by the \bx--space methods~\cite{Breuer09}. The Monte--Carlo method is based on repeatedly corrupting the acquired data with synthetic noise properly scaled and correlated, and assessing the empirical statistics of the GRAPPA reconstructions resulting from these noisy data. Unfortunately, this method is inherently time consuming  and is only  able to provide a noisy estimate of the noise--maps due to the limited number of Monte--Carlo realizations. In contrast, \bx--space methods reformulate the GRAPPA reconstruction as a pixel-wise multiplication in \bx--space, providing computationally efficient noise characterization by avoiding the extensive \bk--space correlations. However, as shown in Refs. \cite{Brau08,Beatty08}, the \bx--space reconstruction is not exactly equivalent to a GRAPPA reconstruction, rendering systematic errors in the estimated noise--maps \cite{Rabanillo17}. Further, errors in  \bx--space methods are generally associated with the boundaries between \bk--space regions with different different sampling patterns~\cite{Rabanillo17}, which is troublesome in 3D acquisitions obtained with highly non--uniform sub-sampling patterns.

In this work, we propose a method for exact noise characterization  for 3D GRAPPA reconstructions under the assumptions of stationarity and uncorrelation in the original \bk--space sub-sampled acquisition. The proposed method extends a recently reported 2D approach \cite{Rabanillo17} to the more challenging analysis of 3D imaging, presenting an exact and computationally efficient solution by exploting the extensive symmetries and separability in each step of the reconstruction. We demonstrate the accuracy of the proposed method under multiple sub-sampling scenarios, including various sub-sampling patterns (Rectangular-type, CAIPIRINHA-type or Random), autocalibration region shapes (rectangular or ellipsoidal) and sampling density (uniform or variable across the \bk--space), on both synthetic and real phantoms. Finally, an \emph{in--vivo} experiment is performed to illustrate the applicability of the proposed noise characterization method.

%%%%%%%%%%%%%%%%%%%%%%%%%%%%%%%%%%%%%%%%
%THEORY
%%%%%%%%%%%%%%%%%%%%%%%%%%%%%%%%%%%%%%%

\section{Theory}
\label{sec:theory}

%GRAPPA----------------------------------------------
\subsection{{Overview of the problem}}
\label{sec:overview}

Parallel MRI techniques achieve acceleration by sub-sampling the  \bk--space data acquisition. In 3D Cartesian acquisitions with parallel imaging, the 3D data-set is sub-sampled along two phase encoding directions in order to exploit the sensitivity variations of the receiver coils in these two spatial dimensions. The acquired data across the $L$ coils ${\vec{{\bf s}} (\boldsymbol k)}=[ s_1 (\boldsymbol k) \cdots s_L (\boldsymbol k) ]$ can be modeled as in \cite{AjaNoiseBook,Thunberg07}: 
\begin{equation} \label{eq:kspaceNoise}
\vec{{\bf s}}(\boldsymbol{k}) = \vec{{\bf a}}(\boldsymbol{k}) + \vec{{\bf n}}(\boldsymbol{k};{\bf \Gamma}_{a},{\bf C}_{a}),
\end{equation}
where ${\vec{{\bf a}} (\boldsymbol k)}=[ a_1(\boldsymbol k) \cdots a_L(\boldsymbol k) ]$ is the noiseless signal in the \bk--space and ${\vec{{\bf n}}(\boldsymbol{k};{\bf \Gamma}_{a},{\bf C}_{a})}$ is the acquisition noise, which is assumed to follow a $L$-variate complex zero-mean normal distribution~(see \cite{Goodman63}) for each \bk--space point. This noise distribution is fully characterized by its covariance matrices ${{\bf \Gamma}_{a}}=E\{\vec{{\bf n}} \cdot \vec{{\bf n}}^H\}$ and ${{\bf C}_{a}}=E\{\vec{{\bf n}} \cdot \vec{{\bf n}}^T\}$, where the operators $^T$ and $^H$ denote the transpose and Hermitian operator of a matrix, respectively. Furthermore, we also assume the acquisition noise in MRI is stationary \cite{AjaNoiseBook}, i.e. ${{\bf \Gamma}_{a}}$ and ${{\bf C}_{a}}$ do not depend on ${\boldsymbol k}$.

GRAPPA reconstructs the \bk--space for each coil by filling the missing points with a linear combination of the acquired points within its neighborhood ${\eta (\boldsymbol k)}$ across all the coils~\cite{Griswold02}:
\begin{equation} \label{eq:GRAPPAeq_k}
s_l^R(\boldsymbol k)=\sum_{m=1}^{L} \sum_{\boldsymbol c \in \eta (\boldsymbol k)} s_m^S (\boldsymbol c) \cdot \omega^{\boldsymbol k}_m(l,\boldsymbol c),
\end{equation}
where $s_m^S({\boldsymbol k})$ is the sampled \bk--space signal from coil $m$, ${\omega^{\boldsymbol k}_m(l,\boldsymbol c)}$ are the complex weights for coil $l$ in the reconstruction kernel, and $s_l^R({\boldsymbol k})$ is the reconstructed \bk--space signal for coil $l$. The weights ${\omega^{\boldsymbol k}_m(l,\boldsymbol c)}$  can be estimated from the data in a fully sampled low-frequency region of the \bk--space, called the Auto Calibration Signal (ACS) lines~(\cite{Griswold02}), or from a separate calibration scan \cite{Setsompop12}. Note the dependence  ${\omega^{\boldsymbol k}_m(l,\boldsymbol c)}$ on the  \bk--space location for the general case in which multiple kernels are used (e.g. for acquisitions with non-uniform sub-sampling patterns).

In order to obtain the final image $S_T(\boldsymbol x)$, it is necessary to combine the data from all the channels $S_l^R(\boldsymbol x)$ obtained by Fourier transformation of  $s_l^R(\boldsymbol x)$. This can be done using the sum of squares (SoS) as in~\cite{AjaMRM11}, or using a properly weighted linear combination as described by \cite{Roemer90}. In this work we used the linear combination proposed in \cite{Walsh00}, where the final image can be expressed as:
\begin{equation} \label{eq:SMF}
S_T(\boldsymbol x)=\vec{{\bf m}}(\boldsymbol x) \cdot \vec{{\bf S}^R} (\boldsymbol x) = \sum_{l=1}^L m_l(\boldsymbol x) \cdot S_l^R(\boldsymbol x),
\end{equation}
where $\vec{{\bf m}} (\boldsymbol x)=\left[m_1 (\boldsymbol x) \cdots m_L(\boldsymbol x) \right]^T$ is a vector combining the information from each coil, and the $\boldsymbol x$ dependence indicates that the operation is pixel--wise.

%% DHA: In equation 4, should we include the dependence of the SNRs and the sigmas on location x? 

The objective in this paper is to quantify the noise amplification associated with the GRAPPA reconstruction in 3D. This noise amplification is typically described in terms of the so-called $g$-factor~(\cite{Breuer09}):
\begin{equation} \label{eq:$g$-factor_def}
g_T(\boldsymbol x) =\frac{\text{SNR}_\text{full}(\boldsymbol x)}{\text{SNR}_\text{acc}(\boldsymbol x)\cdot \sqrt{R_{\text{eff}}}}= \frac{\sigma_\text{acc}(\boldsymbol x)}{\sigma_\text{full} (\boldsymbol x)\cdot \sqrt{R_{\text{eff}}}},
\end{equation} 
where $\text{SNR}_\text{full,acc}$ and $\sigma_\text{full,acc}$ denote the signal-to-noise ratio and the standard deviation in the fully sampled image and in the sub-sampled image after reconstruction, respectively, and $R_{\text{eff}}$ denotes the effective acceleration (ratio between the number of lines in the reconstructed \bk--space and the number of acquired lines). 

\subsection{{\textbf{k}--space method for noise characterization in GRAPPA}}
\label{sec:grappa_k}

Similarly to the previously described 2D imaging case in Ref~\cite{Rabanillo17}, we rely on the following assumptions: (1) GRAPPA weights are non-stochastic, i.e., they are independent of the noise realization. This holds strictly true for the case when they are computed from a separate calibration scan as in \cite{Setsompop12}. Nevertheless, due to the overdetermination of GRAPPA weights estimation, this assumption is common even in self-calibrated acquisitions; (2) noise at different acquired points in the \bk--space is IID \footnote{If the noise shows temporal correlation along the frequency--encoding direction, our analysis can still be applied. This potential correlation would simply increase the width of the neighborhood a reference point correlates with, as described in subsection ``\bk--space Interpolation''.} as in~\cite{Henkelman85,McVeigh85,Kellman05}. 

In order to characterize the noise in the final composite image, we proceed to characterize the noise propagation through the reconstruction step by step (see Fig.~\ref{fig:schemegraph}):

\subsubsection{{\textbf{k}--Space interpolation}}
\label{sec:grappa_interpolation} 
%SPACE INTERPOLATION----------
The reconstruction of a missing point $\vec{\bf s}^R(\boldsymbol k)$ can be  expressed by rewriting Eq.~\eqref{eq:GRAPPAeq_k} to a matrix operation:
\begin{equation} \label{eq:GrappaRecPoint}
\vec{\bf s}^R(\boldsymbol k)=\sum_{{\bf c} \in \eta (\boldsymbol k)} \boldsymbol{\W}^{\boldsymbol k}_{{\bf c}}  \cdot \vec{\bf s}^S({\bf c}),
\end{equation}
where the vector ${\vec{\bf s}^S({\bf c})=\left[ s_1^S({\bf c}) \cdots s_L^S({\bf c}) \right]}$ contains the \bk--space acquired data within the neighborhood ${\eta(\boldsymbol k)}$. Also, $\boldsymbol{\W}^{\boldsymbol k}_{{\bf c}}$ is an ${L\times L}$ matrix in which the $l$-th row contains the GRAPPA weights ${\vec{\boldsymbol{\omega}}^{\boldsymbol k}(l,{\bf c}) = [\omega^{\boldsymbol k}_1(l,{\bf c})	\ldots \omega^{\boldsymbol k}_L(l,{\bf c})]}$ 
from Eq.~\eqref{eq:GRAPPAeq_k} associated to location ${\bf c}$ to reconstruct the $l$-th coil element of the \bk--space vector $\vec{\bf s}^R(\boldsymbol k)$. Since we consider the general case of using multiple kernels, this matrix can change through \bk--space. Consequently, $\boldsymbol{\W}^{\boldsymbol k}_{{\bf c}}$ depends on both the point reconstructed $\boldsymbol k$ and the point considered for the reconstruction ${\bf c}$.

Based on this reconstruction, the acquisition noise $\vec{\bf n}^S({\boldsymbol k})$ will propagate into the reconstructed \bk--space as follows:
\begin{equation} \label{eq:GrappaRecPoint}
\vec{\bf n}^R(\boldsymbol k)=\sum_{{\bf c} \in \eta (\boldsymbol k)} \boldsymbol{\W}^{\boldsymbol k}_{{\bf c}}  \cdot \vec{\bf n}^S({\bf c}),
\end{equation}
where the vectors ${\vec{\bf n}^S({\bf c})}$ and ${\vec{\bf n}^R({\bf c})}$ are the noise in the acquired and reconstructed \bk--space, respectively. 

Further, the GRAPPA interpolation introduces correlation between two arbitrary points in the reconstructed \bk-space. After this first stage of the reconstruction, the resulting noise correlation matrices ${\bf \Gamma}_1$ and ${\bf C}_1$ (where the index $1$ simply indicates the order of appearance of these correlation matrices in the present derivation) can be expressed as
\begin{equation} \label{eq:GammaGrappa}
\begin{split}
{\bf \Gamma}_1^{ij} & = E\{\vec{\bf n}^R(\boldsymbol k_i) \cdot \vec{\bf n}^R(\boldsymbol k_j) ^H\} \\
& = E\left\{\left(\sum_{{\bf c}_i \in \eta_i} \boldsymbol{\W}^{\boldsymbol k_i}_{{\bf c}_i}  \cdot \vec{\bf s}^S({\bf c}_i) \right) \left(\sum_{{\bf c}_j \in \eta_j} \boldsymbol{\W}^{\boldsymbol k_j}_{{\bf c}_j}  \cdot \vec{\bf s}^S({\bf c}_j) \right)^H\right\} \\
& = \sum_{{\bf c}_i \in \eta_i} \sum_{{\bf c}_j \in \eta_j} \boldsymbol{\W}^{\boldsymbol k_i}_{{\bf c}_i} \cdot E\left\{ \vec{\bf s}^S({\bf c}_i) \cdot \vec{\bf s}^S({\bf c}_j)^H\right\} \cdot {\boldsymbol{\W}^{\boldsymbol k_j}_{{\bf c}_j}}^H \\
& = \sum_{{\bf c} \in \left( \eta_i \cap \eta_j \right)}  \boldsymbol{\W}^{\boldsymbol k_i}_{{\bf c}_i} \cdot {\bf \Gamma}_a \cdot {\boldsymbol{\W}^{\boldsymbol k_i}_{{\bf c}_j}}^H  \\
\end{split}
\end{equation}
and equivalently
\begin{equation} \label{eq:CGrappa}
{\bf C}_1^{ij}=E\{\vec{\bf n}^R(\boldsymbol k_i) \cdot \vec{\bf n}^R(\boldsymbol k_j) ^T\}=\sum_{{\bf c} \in \left( \eta_i \cap \eta_j \right)}  \boldsymbol{\W}^{\boldsymbol k_i}_{{\bf c}_i} \cdot {\bf C}_a \cdot {\boldsymbol{\W}^{\boldsymbol k_i}_{{\bf c}_j}}^T.
\end{equation}
As a result, a point will be correlated to any point within overlapping neighborhoods. We will reconstruct a \bk--space of size $\left[N_{p_1},N_{p_2},N_f\right]$, where $\{ p_1,p_2 \}$ denote the two phase-encoding direction and $f$ denotes the frequency-encoding direction. We will hereafter refer to a column as the data along the first dimension (fixing the other two dimensions), a row the data along the second dimension and a layer the data along the third dimension. Furthermore, from now on we focus on the calculation of matrix $\Gamma$ since the computation of the final matrix $C$ is analogous.

Eqs.~\eqref{eq:GammaGrappa}--\eqref{eq:CGrappa} allow to compute the correlation between any two points in \bk--space. Our objective is to obtain the correlation matrices between all the points that correlate with a reference column prior to considering the effect of an iFFT along the column dimension, in order to keep track of all correlations when this column iFFT (next subsection) is performed. If we define the size of the reconstruction kernel as ${ \left[K_{p_1},K_{p_2},K_f\right]}$, a reconstructed point will correlate with ${2K_f-1}$ points in the frequency-encoding direction. Importantly, its correlations along the phase-encoding directions depend on the layer since we may use different kernels across the \bk--space. From now on, we will use the term reconstruction pattern of a column to refer to the set containing the position of the points that are used in its reconstruction, as well as the kernel used for each of them. Taking into account that all the columns (or rows) behave equally across the third dimension, and in order to operate efficiently, our algorithm starts by identifying the columns that present a unique reconstruction pattern with respect to its neighboring columns, i.e., all the columns with the same reconstruction pattern will have the same correlation matrix. In this way we will avoid the unnecessary burden of computing the same correlation matrices multiple times.

Let $N_{uq}$ be the number of columns with unique reconstruction patterns. For each of these columns with pattern $m_{uq}$ (where $\{m_{uq} = 1,\ldots ,N_{uq}\}$), we can compute its correlation matrix ${\bf \Gamma}_2^{m_{uq}}$ with its neighboring columns:
\begin{equation} \label{eq:GammaInter}
{\bf \Gamma}_2^{m_{uq}} = \left( \begin{array}{cccc}
{\bf B}_{11}^{m_{uq}} & {\bf B}_{12}^{m_{uq}} & \cdots & {\bf B}_{1L}^{m_{uq}}\\
{\bf B}_{21}^{m_{uq}} & {\bf B}_{22}^{m_{uq}} & \cdots & {\bf B}_{2L}^{m_{uq}}\\
\vdots & \vdots & \ddots & \vdots\\
{\bf B}_{L1}^{m_{uq}} & {\bf B}_{L2}^{m_{uq}} & \cdots & {\bf B}_{LL}^{m_{uq}}
\end{array}\right),
\end{equation}
where ${\bf B}_{ij}^{m_{uq}}$ contains the correlations of the (say) reference column and its neighboring columns in the $i$-th coil and these equivalent columns in the $j$-th coil. Due to its Hermitian symmetry ${\bf B}_{ij}^{m_{uq}}=({\bf B}_{ji}^{m_{uq}})^H$ we only need to compute ${L\cdot (L+1)/2}$ blocks. 

Each matrix ${\bf B}_{ij}^{m_{uq}}$ is equally composed of sub-blocks  of size $N_{p_1}\times N_{p_1}$ containing the correlation between each pair of the columns contained in the neighborhood. Since the iFFT along the first dimension operates on each column independently, these sub-blocks will be transformed independently as we will show in the following sections. Furthermore, it is important to notice the stationarity along the third dimension (see \cite{Rabanillo17}) as well as the fact that we repeat this computation for each column with a unique pattern. This implies that we only need to compute the central row of the block matrix ${\bf B}_{ij}^{m_{uq}}$ containing the correlation sub-blocks $\{ {\bf b}^{m_{uq}}_{ij,m_{uq}^{p_2}}, m_{uq}^{p_2} =1,\ldots, M_{uq}^{p_2} \}$ between the reference column and its neighboring columns. 

The total number of columns considered for a reference column is given by $M_{uq}=(2K_f-1)\cdot M_{uq}^{p_2}$, where $M_{uq}^{p_2}$ is the number of columns that the reference column correlates with in the second dimension, which depends on the pattern indexed by $m_{uq}$. Therefore, the overall number of sub-blocks we need to keep track of is $M_{Tot}=\sum_{m_{uq} =1}^{M_{uq}}{M_{uq}^{p_2} \cdot (2K_f-1) \cdot L\cdot (L+1)/2}$.

%COLUMN IFFT------------------------------

\subsubsection{{Column iFFT}}
\label{sec:grappa_col_iFFT} 

After filling the missing lines in the \bk--space with GRAPPA interpolation, the data are transformed into the image--space through a 3D iFFT, which is equivalent to successively computing a 1D-iFFT along each of the dimensions. If we start by applying the iFFT along the first dimension, the number of columns that a reference column correlates with is preserved. Furthermore, since the 1D-iFFT can be expressed as a matrix product using the Fourier associated matrix ${{\bf F}_I}$, the noise propagates using the following block-diagonal matrix
\begin{equation} \label{eq:ComposeiFFT}
\boldsymbol{\F}_I = \left( \begin{array}{cccc}
{\bf F}_I & \bf \overline{0} & \cdots & \bf \overline{0}\\
\bf \overline{0} & {\bf F}_I & \cdots & \bf \overline{0}\\
\vdots & \vdots & \ddots & \vdots\\
\bf \overline{0} & \bf \overline{0} & \cdots & {\bf F}_I
\end{array}\right).
\end{equation}
with $M_{uq} \cdot L$ blocks, which in the hybrid--space $\left[ x,k_{p_2},k_f\right]$ results in the correlation matrix
\begin{equation} \label{eq:GammaHybridCol}
{\bf \Gamma}_3^{m_{uq}} = \boldsymbol{\F}_I \cdot {\bf \Gamma}_2^{m_{uq}} \cdot \boldsymbol{\F}_I^H = \left( \begin{array}{cccc}
{\bf D}_{11}^{m_{uq}} & {\bf D}_{12}^{m_{uq}} & \cdots & {\bf D}_{1L}^{m_{uq}}\\
{\bf D}_{21}^{m_{uq}} & {\bf D}_{22}^{m_{uq}} & \cdots & {\bf D}_{2L}^{m_{uq}}\\
\vdots & \vdots & \ddots & \vdots\\
{\bf D}_{L1}^{m_{uq}} & {\bf D}_{L2}^{m_{uq}} & \cdots & {\bf D}_{LL}^{m_{uq}}
\end{array}\right),
\end{equation}
where each ${\bf D}_{ij}^{m_{uq}}$ block is equally composed of sub--blocks $\{ {\bf d}^{m_{uq}}_{ij,m_{uq}^{p_2}}, m_{uq}^{p_2} =1,\ldots, M_{uq}^{p_2} \}$ that can be computed independently by:
\begin{equation} \label{eq:FFTblockCol}
{\bf d}^{m_{uq}}_{ij,m_{uq}^{p_2}} = {\bf F}_I \cdot {\bf b}^{m_{uq}}_{ij,m_{uq}^{p_2}} \cdot {\bf F}_I^H.
\end{equation}

Importantly, columns whose reconstruction pattern is a shifted version (along the first dimension) of another column are considered to have an equivalent (non-unique) pattern, thereby reducing the computational requirements for the proposed algorithm. This is due to the fact that the diagonal of ${\bf \Gamma}_3^{m_{uq}}$ for two columns with equivalent pattern (except for a shift) is equal (see Appendix) and, as shown in the following section, the proposed algorithm only relies on the the diagonal of these matrices at this step.

%ROW IFFT---------------------------------------

\subsubsection{{Row iFFT}}
\label{sec:grappa_row_iFFT}

The next step in the 3D-iFFT is to perform the 1D-iFFT along the second (row) dimension. At this step, it suffices to keep track of the correlations between the points with the same coordinate in the first dimension, since in the following stages points that are not in this first coordinate will not be combined. 

Additionally, all the rows along the third (layer) dimension behave equivalently due to the stationarity of the GRAPPA reconstruction along the layer dimension. For each row along the first dimension (dependence on $k_f$ is removed due to the aforementioned stationarity along the layers), as shown with the $x$-dependence, we will create the correlation matrix:
\begin{equation} \label{eq:GammaHybridRow}
{\bf \Gamma}_4 (x) = \left( \begin{array}{cccc}
{\bf G}_{11} & {\bf G}_{12} & \cdots & {\bf G}_{1L}\\
{\bf G}_{21} & {\bf G}_{22} & \cdots & {\bf G}_{2L}\\
\vdots & \vdots & \ddots & \vdots\\
{\bf G}_{L1} & {\bf G}_{L2} & \cdots & {\bf G}_{LL}
\end{array}\right),
\end{equation}
where each block ${\bf G}_{lm}$ contains the correlations between the reference row and the $2K_f-1$ adjacent rows along the third dimension in coils $l$ and $m$ within the hybrid--space ${(x,k_{p_2},k_f)}$. Due to the stationarity of the GRAPPA kernel along the frequency-encoding dimension and the assumption of noise stationarity, the correlation between a reference row and its surrounding rows is the same, independently of the row picked for reference, giving rise to a block-Toeplitz structure. As in \cite{Rabanillo17}, if we index the $2K_f-1$ rows placing the reference as the central one (index $K_f$), then the first column of the block ${\bf G}_{ij}$ (with length $2K_f-1$ blocks) is defined as
\begin{equation} \label{eq:FirstColumnOfG}
\left [\begin{array}{ccccccc}
{\bf g}^{0}_{ij} & {\bf g}^{1}_{ij} & \cdots & {\bf g}^{K_f-1}_{ij} & \bf \overline{0} & \cdots & \bf \overline{0} 
\end{array} \right ]^T,
\end{equation}
where ${\bf g}^{m}_{ij}$ is the $N_{p_2}\times N_{p_2}$ covariance matrix of two columns, the subtraction of the indices of which equals $m$, and $\bf \overline{0}$ denotes a null matrix of dimensions $N_{p_2}\times N_{p_2}$. These correlations are obtained by simply reallocating the elements of the diagonal ---at position $x$--- from the previously computed blocks ${\bf d}^{m_{uq}}_{ij,m_{uq}^{p_2}}$ for every unique column in Eq.~\eqref{eq:FFTblockCol}. As for the first row, we get
\begin{equation} \label{eq:FirstRowOfG}
\left [\begin{array}{ccccccc}
{\bf g}^0_{ij} & {\bf g}^{-1}_{ij} & \cdots & {\bf g}^{1-K_f}_{ij} & \bf \overline{0} & \cdots & \bf \overline{0} 
\end{array} \right ].
\end{equation}

Due to the block-Toeplitz structure and the presence of null correlations known beforehand, we only need to compute ${2K_f-1}$ sub-blocks. Therefore, the overall number of sub-blocks needed to build Eq.\eqref{eq:GammaHybridRow} is at most ${(2K_f-1) \cdot L\cdot (L+1)/2}$. Finally, computing the row 1D-iFFT provides a correlation matrix given by:
\begin{equation} \label{eq:GammaFFTRow}
{\bf \Gamma}_5 (x) = \boldsymbol{\F}_I \cdot {\bf \Gamma}_4 (x) \cdot \boldsymbol{\F}_I^H = \left( \begin{array}{cccc}
{\bf H}_{11} & {\bf H}_{12} & \cdots & {\bf H}_{1L}\\
{\bf H}_{21} & {\bf H}_{22} & \cdots & {\bf H}_{2L}\\
\vdots & \vdots & \ddots & \vdots\\
{\bf H}_{L1} & {\bf H}_{L2} & \cdots & {\bf H}_{LL}
\end{array}\right),
\end{equation}
where the blocks ${\bf H}_{ij}$ are in turn composed of sub-blocks ${\bf h}^{m}_{ij}$ that are obtained by:
\begin{equation} \label{eq:FFTblockRow}
{\bf h}^{m}_{ij} = {\bf F_I} \cdot {\bf g}^{m}_{ij} \cdot {\bf F_I}^H.
\end{equation}

\subsubsection{{Layer iFFT}}
\label{sec:grappa_layer_iFFT}

The last step to complete the 3D-iFFT is the 1D-iFFT along the third dimension, for which we only need the correlation within each layer. For each layer, we can build the matrix containing the correlation within that layer accross all the coil channels:
\begin{equation} \label{eq:GammaHybridLayer}
{\bf \Gamma}_6 (x,y) = \left( \begin{array}{cccc}
{\bf U}_{11} & {\bf U}_{12} & \cdots & {\bf U}_{1L}\\
{\bf U}_{21} & {\bf U}_{22} & \cdots & {\bf U}_{2L}\\
\vdots & \vdots & \ddots & \vdots\\
{\bf U}_{L1} & {\bf U}_{L2} & \cdots & {\bf U}_{LL}
\end{array}\right),
\end{equation}
where every block ${\bf U}_{lm}$ has dimensions ${N_f \times N_f}$ and contains the (cross-)correlations of the selected layer in coils $l$ and $m$ within the hybrid space ${(x,y,k_f)}$. These correlations are obtained by selecting the appropriate components in Eq.~\eqref{eq:GammaFFTRow}. Since GRAPPA performs a circular interpolation, the reconstruction shows a cyclical structure along the frequency-encoding direction. We define the first column of block ${\bf U}_{ij}$ as
\begin{equation} \label{eq:FistRowUij}
\vec{{\bf u}_{ij}}=\left [ u^0_{ij}, \  u^{1}_{ij}, \  \cdots, \  u^{K_f-1}_{ij}, \  0, \ \cdots, \  0, \   u^{1-K_f}_{ij}, \  \cdots, \   u^{-1}_{ij}\right]^T,
\end{equation}
with $N_f-(2K_f-1)$ zeroes and values $u^{m}_{ij}$ taken from the components in sub-block ${\bf h}^{m}_{ij}$ in Eq.\eqref{eq:FFTblockCol} that correspond to the selected layer. As a result of the stationarity along the third dimension, the $j$-th row of ${\bf U}_{ij}$, $2\leq j \leq N_f-1$, is obtained as a simple rightward circular shift of the row $j-1$, which results in the blocks ${\bf U}_{lm}$ being circulant matrices. The correlation matrix after the 1D-iFFT is given by:
\begin{equation} \label{eq:GammaCoilsImage}
{\bf \Gamma}_7 (x,y) = \boldsymbol{\F}_I \cdot {\bf \Gamma}_6 (x,y) \cdot \boldsymbol{\F}_I^H = \left( \begin{array}{cccc}
{\bf V}_{11} & {\bf V}_{12} & \cdots & {\bf V}_{1L}\\
{\bf V}_{21} & {\bf V}_{22} & \cdots & {\bf V}_{2L}\\
\vdots & \vdots & \ddots & \vdots\\
{\bf V}_{L1} & {\bf V}_{L2} & \cdots & {\bf V}_{LL}
\end{array}\right),
\end{equation}
where the blocks ${\bf V}_{ij}$ are computed as:
\begin{equation} \label{eq:FFTblockLayer}
{\bf V}_{ij} = {\bf F_I} \cdot {\bf U}_{ij} \cdot {\bf F_I}^H.
\end{equation}

Finally, due to the fact that circulant matrices are diagonalized by the FT~(see \cite{DavisBook}), Eq.~\eqref{eq:FFTblockLayer} simplifies to:
\begin{equation} \label{eq:FFTblockLayerCirculant}
{\bf V}_{ij} = {\bf F}_I \cdot {\bf U}_{ij} \cdot {\bf F}_I^H = \text{diag}({\bf F}_I \cdot \vec{{\bf u}_{ij} }).
\end{equation}

Note that we only need to compute ${\left[ N_f \cdot L \cdot (L+1)/2 \right]}$ 1D-iFFT due to Hermitian symmetry in this step.

%COIL COMBINATION--------------------------------
\subsubsection{{Coil combination}}
\label{sec:coil_combination}

In GRAPPA reconstructions, the  images for each of the individual channels are combined to obtain the coil-combined `composite' image. If the SoS is used  for coil combination, as in~\cite{AjaMRM11}, the noise distribution in the composite image is well approximated by a non--central--$\chi$ distribution. Alternatively, if a linear combination is used as suggested in \cite{Walsh00} (see Eq.~\eqref{eq:SMF}), the Gaussian behaviour is preserved in the composite image (the distribution becomes Rician if the magnitude is taken). In order to characterize the distribution at every pixel in the final image, we need the correlation matrices for each pixel $\vec{\bf p}$ along the coil dimensions, which are contained in ${\bf \Gamma}_7$. Since the 1D-iFFT in the frequency-encoding direction introduces non-stationarity along this dimension, we proceed pixel by pixel and extract the correlation matrices ${\bf \Gamma}_8$ and ${\bf C}_8$ for a selected position, which consists of picking the corresponding entry in the diagonal of ${\bf \Gamma}_7$ or ${\bf C}_7$. Applying a linear combination as in Eq.~\eqref{eq:SMF} preserves the Gaussianity in the composite image, although this image will show spatial correlations and non-stationarity. We proceed as in \cite{Rabanillo17} and for every pixel in the final image we define

\begin{equation} \label{eq:GammaSMF}
{\Gamma}_9(\boldsymbol x)=\vec{{\bf m}}_l(\boldsymbol x) \cdot {\bf \Gamma}_8(\boldsymbol x) \cdot \vec{{\bf m}}_l^H(\boldsymbol x),
\end{equation}
then the variance of noise for the real and imaginary components can be calculated by:
\begin{equation} \label{eqn:finalSigma}
\begin{split}
\sigma^2_{\text{re},{\mathcal{R}}}(\boldsymbol x)=\frac{1}{2} \text{Re}\left\lbrace {\Gamma}_9 + {C}_9 \right\rbrace,  \\
\sigma^2_{\text{im},{\mathcal{R}}}(\boldsymbol x)=\frac{1}{2} \text{Re}\left\lbrace {\Gamma}_9 - {C}_9 \right\rbrace,
\end{split}
\end{equation}
where $\mathcal{R}$ indicates that this is the noise in the reconstructed sub-sampled image. Note that correlation between real and imaginary components for a pixel can exist, which are computed as:
\begin{equation} \label{eqn:finalSigmaRealImag}
\begin{split}
\sigma_{\text{re-im},\mathcal{R}}^2(\boldsymbol x)=\frac{1}{2} \text{Im}\left\lbrace C_7 - \Gamma_7 \right\rbrace,  \\
\sigma_{\text{im-re},\mathcal{R}}^2(\boldsymbol x)=\frac{1}{2} \text{Im}\left\lbrace \Gamma_7 + C_7 \right\rbrace.	
\end{split}
\end{equation}

Finally, the $g$-factor map can be derived from the previous equation defining an {\em average} variance ${\sigma_{\mathcal{R}}^2(\boldsymbol x)}$
\begin{equation} \label{eqn:meanVar}
\sigma_{\mathcal{R}}^2(\boldsymbol x)=\frac{ \sigma_{\text{re},\mathcal{R}}^2(\boldsymbol x) + \sigma_{\text{im},\mathcal{R}}^2(\boldsymbol x)}{2},
\end{equation}
\begin{equation} \label{eq:$g$-factor_final}
g_{\mathcal{R}}(\boldsymbol x)=\frac{\sqrt{\sigma_{\text{re},\mathcal{R}}^2(\boldsymbol x) + \sigma_{\text{im},\mathcal{R}}^2(\boldsymbol x)}}{\sqrt{R_{\text{eff}}} \cdot \sqrt{\vec{{\bf m}}(\boldsymbol x) \cdot \left( {\bf \Sigma}^{\text{full}}_{x,\text{re}} + {\bf \Sigma}^{\text{full}}_{x,\text{im}} \right) \cdot \vec{{\bf m}}^H(\boldsymbol x)}},
\end{equation}
where ${\bf \Sigma}^{\text{full}}_{x,\text{re}}$ and ${\bf \Sigma}^{\text{full}}_{x,\text{im}}$ refer to the covariance matrices for the real and imaginary parts in the coil images when the \bk--space is fully sampled.

%-------------------------------
\subsubsection{{Summary of the procedure}}

The procedure to obtain ${\bf \Gamma}$ matrices is graphically depicted in Figure \ref{fig:schemegraph} and  can be summarized as follows ($\bf C$ matrices are obtained similarly): 
\begin{enumerate}

\item Identify the number of columns with a unique sampling pattern. Columns whose pattern matches a shifted version of another column are not explicitly considered.

\item Calculate the sub-blocks ${\bf b}^{m_{uq}}_{ij,m_{uq}^{p_2}}$ using Eq.~\eqref{eq:GammaGrappa}. The number of such sub-blocks is $M_{Tot}=\sum_{m_{uq} =1}^{M_{uq}}{M_{uq}^{p_2} \cdot (2K_f-1) \cdot L\cdot (L+1)/2}$, and their dimension is $N_{p_1}\times N_{p_1}$. Each entry in these sub-blocks is obtained from Eq.~\eqref{eq:GammaGrappa}, by choosing the appropriate components. 

\item Calculate the sub-blocks ${\bf d}^{m_{uq}}_{ij,m_{uq}^{p_2}}$ resulting from the column 1D-iFFT  using Eq.~\eqref{eq:FFTblockCol}.

\item For each row, build the sub-blocks ${\bf g}^{m}_{ij}$ that build up the  matrix Eq.~\eqref{eq:GammaHybridRow} by properly reallocating the elements of the diagonal from ${\bf d}^{m_{uq}}_{ij,m_{uq}^{p_2}}$.

\item Compute the sub-blocks ${\bf h}^{m}_{ij}$ resulting from the row 1D-iFFT using Eq.~\eqref{eq:FFTblockRow}.

\item For each row and each layer, calculate $\vec{{\bf u}_{ij}}$ in Eq.~\eqref{eq:FistRowUij} by choosing the appropriate components of ${\bf h}^{m}_{ij}$. 

\item Calculate ${\bf U}_{ij}$ resulting from the layer 1D-iFFT  using Eq.~\eqref{eq:FFTblockLayerCirculant}.

\item For every pixel, create its matrix ${\bf \Gamma}_8$ by selecting the right element in the diagonal of ${\bf U}_{ij}$. 

\item Apply equations Eq.~\eqref{eq:GammaSMF}, Eq.~\eqref{eqn:meanVar} and Eq.~\eqref{eq:$g$-factor_final} to obtain pixel-wise noise characterization. 

\end{enumerate}

\begin{figure}[tb]
\centering
\includegraphics[width=0.5\textwidth]{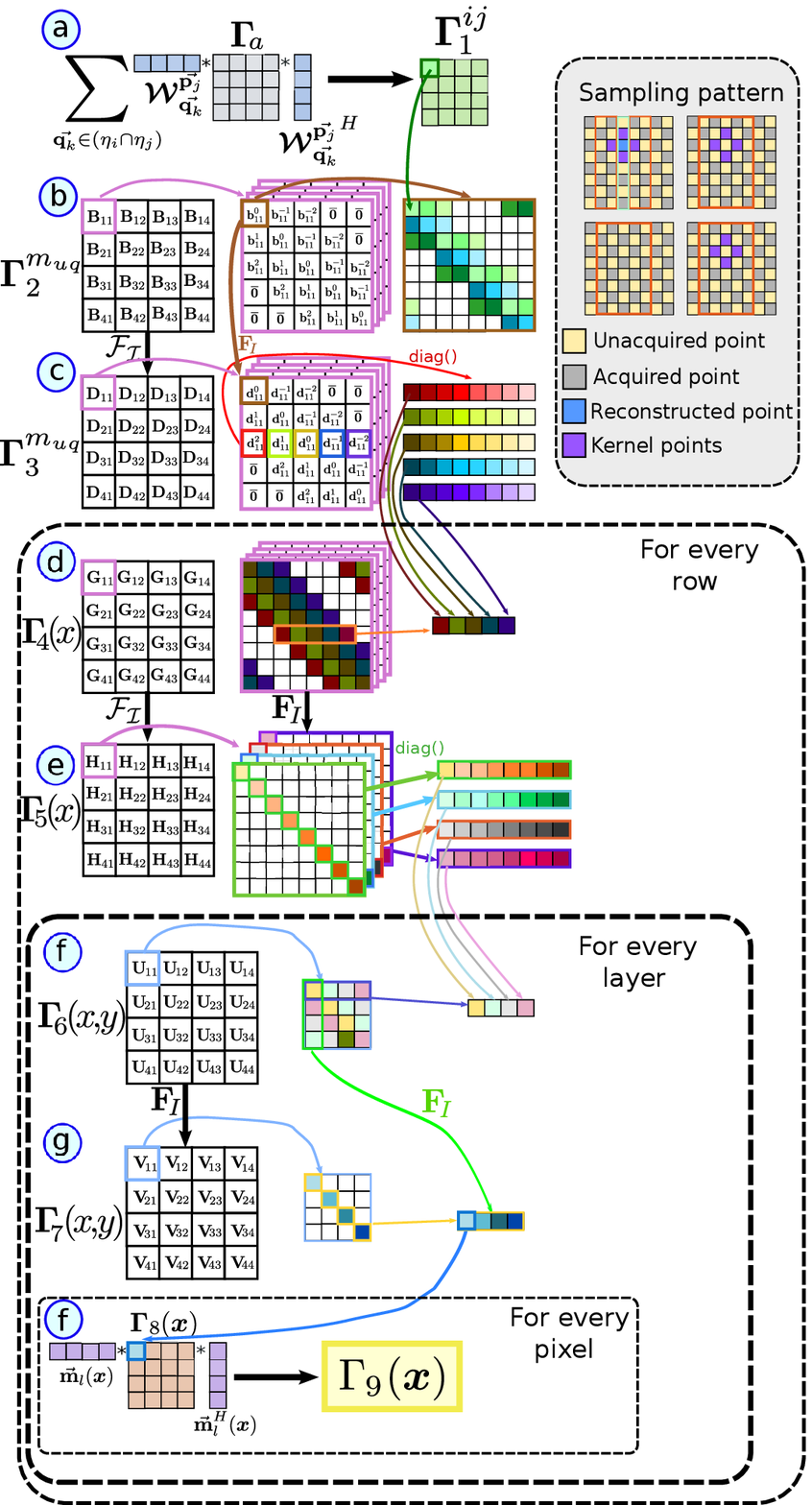}
\caption{Graphical description of the algorithm. For this example we consider a matrix size $8\times 8 \times 4$ with the sub-sampling pattern shown in the upper right corner and L=4 coils. (a) Each reconstructed point (`reference point') is correlated with all the points in its kernel as well as with all the reconstructed points whose kernel overlaps with the kernel of the reference point. In contrast, the acquired points correlate only with the reconstructed points that have them on their kernel. The point to point correlations can be computed using Eq.~\eqref{eq:GammaGrappa}. (b) A reference column (green column) is correlated with $2K_{p_2}-1$ columns along the second phase--encoding dimension, and with all the columns along the frequency--encoding dimension, marked in orange. These correlations are stored in ${\bf \Gamma}_2^{m_{uq}}$, which shows a block structure. (c) Computing the iFFT along the first phase-encoding direction results in ${\bf \Gamma}_3^{m_{uq}}$, where each of the blocks can be computed separately using Eq.~\eqref{eq:FFTblockCol}, and we can keep only the diagonal of the blocks since we only need to keep the correlations between the points that share the position along the first dimension. (d) For every row, we rearrange properly the correlation of a point and its $2K_f-1$ neighboors along all the layers to build ${\bf \Gamma}_4 (x)$. In this particular case, due to the uniform sub-sampling pattern, the correlations are stationary across the rows, but this may vary with other sub-sampling patterns. (e) The iFFT along the second phase-encoding direction that results in ${\bf \Gamma}_5 (x)$ can be computed block by block using Eq.~\eqref{eq:FFTblockRow}, and we can keep only the diagonal of the blocks since we only need to keep the correlations along the layers. (f) These correlations are stationary across the layer, resulting in a Toeplitz structure for every block. (g) The iFFT along the frequency-encoding direction results in diagonal blocks due to this Toeplitz structure and can be computed very efficiently using Eq.~\eqref{eq:FFTblockLayerCirculant}. (h) Finally, for every pixel in the row we simply apply Eq.~\eqref{eq:GammaSMF} using the Walsh coil-combination vector.}
\label{fig:schemegraph}
\end{figure}

%COMPLEXITY------
\subsection{{Computational complexity}}

In terms of memory requirements, the most demanding step in our method is the computation of the correlation matrices between each unique column and its neighbors in the original \bk--space. Since we operate sequentially along the second dimension (i.e., in each iteration of a loop we consider only the columns whose distance to the reference column coincide), we need to store  ${\left[(2K_f-1) \cdot L \cdot (L+1)/2 \cdot M_{uq}\right]}$ complex matrices of size ${ N_{p_1}\times N_{p_1}}$, and after the iFFT we only keep the diagonal of each matrix. However, if memory constraints demand so, it would be possible to reduce the number of matrices stored at the same time by simply operating sequentially along the third dimension or the coil dimension as well. 

In terms of the number of operations, the most demanding step is the computation of the iFFT. It is necessary to compute 
${\left[(2K_f-1) \cdot L \cdot (L+1)/2 \cdot M_{uq} \cdot \bar{M_{uq}^{p_2}} \right]}$ 2D-iFFT for the first phase-encoding direction, where $\bar{M_{uq}^{p_2}}$ denotes the average number of columns that a unique column correlates with along the second dimension; ${\left((2K_f-1) \cdot L \cdot (L+1)/2 \cdot N_{p_1} \right]}$ 2D-iFFT for the second phase-encoding direction and ${\left[N_{p_1} \cdot N_{p_2} \cdot L \cdot (L+1)/2 \right]}$ for the frequency-encoding direction, resulting in a total number of ${(4K_f-1) \cdot \left[ L \cdot (L+1)/2 \right] \cdot N_{p_1} \cdot N_{p_2} \cdot O\left( \overline{N_i \cdot \log(N_i)} \right)}$ complex multiplications approximately, where $\overline{N_i \cdot \log(N_i)}$ refers to $\left( N_{p_1} \cdot \log(N_{p_1}) + N_{p_2} \cdot \log(N_{p_2}) + N_f \cdot \log(N_f) \right)$.

There is an alternative approach to apply our method that consists of following the noise correlation through an alternative GRAPPA reconstruction performed in the hybrid--space ${(k_{p_2},k_{p_2},z)}$ as in \cite{Brau08}. In this reconstruction, a missing point would  only be interpolated from points that share the same $z$-coordinate, and thus there would not be any correlations along the frequency--encoding direction. This would allow to store only ${\left[L \cdot (L+1)/2 \cdot M_{uq}\right]}$ complex matrices of size ${ N_{p_1}\times N_{p_1}}$. However, the reconstruction weights are not stationary anymore through \bk--space, which would require to compute 
${\left[N_f \cdot L \cdot (L+1)/2 \cdot M_{uq} \cdot \bar{M_{uq}^{p_2}} \right]}$ 2D-iFFT for the first phase-encoding direction and ${\left(N_f \cdot L \cdot (L+1)/2 \cdot N_{p_1} \right]}$ 2D-iFFT for the second phase-encoding direction. This would result in a total number of ${N_f \cdot N_{p_1} \cdot N_{p_2} \cdot \left[ L \cdot (L+1)/2 \right] \cdot  O\left( \overline{N_i \cdot \log(N_i)}' \right)}$ complex multiplications approximately, where $\overline{N_i \cdot \log(N_i)}'=\sum_{i=1}^2 N_{p_i} \cdot \log(N_{p_i})$. In summary, this alternative approach would reduce the memory requirements, but at the expense of increasing the number of operations.

%%%%%%%%%%%%%%%%%%%%%%%%%%%%%%%%%%%%%%%%%%%%%%%%%%%%%%%
%METHODS---------------
%%%%%%%%%%%%%%%%%%%%%%%%%%%%%%%%%%%%%%%%%%%%%%%%%%%%%%

\section{Methods} 

Three data sets are considered for the experiments (see Fig.~\ref{fig:datasets})

\begin{itemize}
	\item {\em Simulated abdomen data set:} we have synthetized a 3D volume using the simulation environment XCAT \cite{Segars10} based on the extended cardio-torso phantom. We simulated a T1-weighted acquisition using the following acquisition parameters: TE/TR=1.5/3ms, flip angle=60$^\circ$. A 32-coil acquisition was simulated by modulating the image using artificial sensitivity maps coded for each coil as in \cite{AjaNoiseBook,AjaSENSE}. The noise-free coil images were transformed into the \bk--space and corrupted with synthetic Gaussian noise characterized by the matrices ${\bf \Gamma}_k$ and ${\bf C}_k$ with SNR=25 for each coil, and the correlation coefficient between coils was set to $\rho=0.1$. For statistical purposes, 4000 realizations of each image were used. 
	\item {\em Water phantom acquisition:} A MR phantom sphere with solution (GE Medical Systems, Milwaukee, WI) was scanned in a 32-channel head coil on a 3.0T scanner (MR750, GE Healthcare, Waukesha, WI). A spoiled gradient-echo acquisition with 100 realizations of the same fully-encoded kspace sampling was used. Acquisition parameters included: coronal view, TE/TR=0.96/3.69ms, flip angle=12$^\circ$, field of view=22$\times$22$\times$30.7cm$^3$, acquisition matrix size=60$\times$60$\times$32, bandwidth$=\pm62.5$KHz. We corrected for B$_0$ field drift related phase variations and magnitude decay \cite{Vos16} by a pre-processing step. First we estimated the phase-shift between realizations from the center of the \bk--space as a cubic function of time and removed it afterwards. And second we estimated the magnitude-decay in the \bk--space as a linear function and substracted it in order not to affect the noise. 
	\item {\em In vivo acquisition:} in order to assess the feasibility of the proposed method, after obtaining the approval fo the local institutional review board (IRB), a volunteer was scanned in a 32-channel head coil on a 3.0T scanner (MR750, GE Healthcare, Waukesha, WI). A spoiled gradient-echo acquisition with 100 realizations of the same fully-encoded kspace sampling was used. Acquisition parameters included: coronal view, TE/TR=0.83/3.69ms, field of view=22$\times$22$\times$21.6cm$^3$, matrix size=64$\times$64$\times$36, bandwidth$=\pm62.5$KHz. 
\end{itemize}

\begin{figure}[tb]
\centering
\includegraphics[width=0.99\columnwidth]{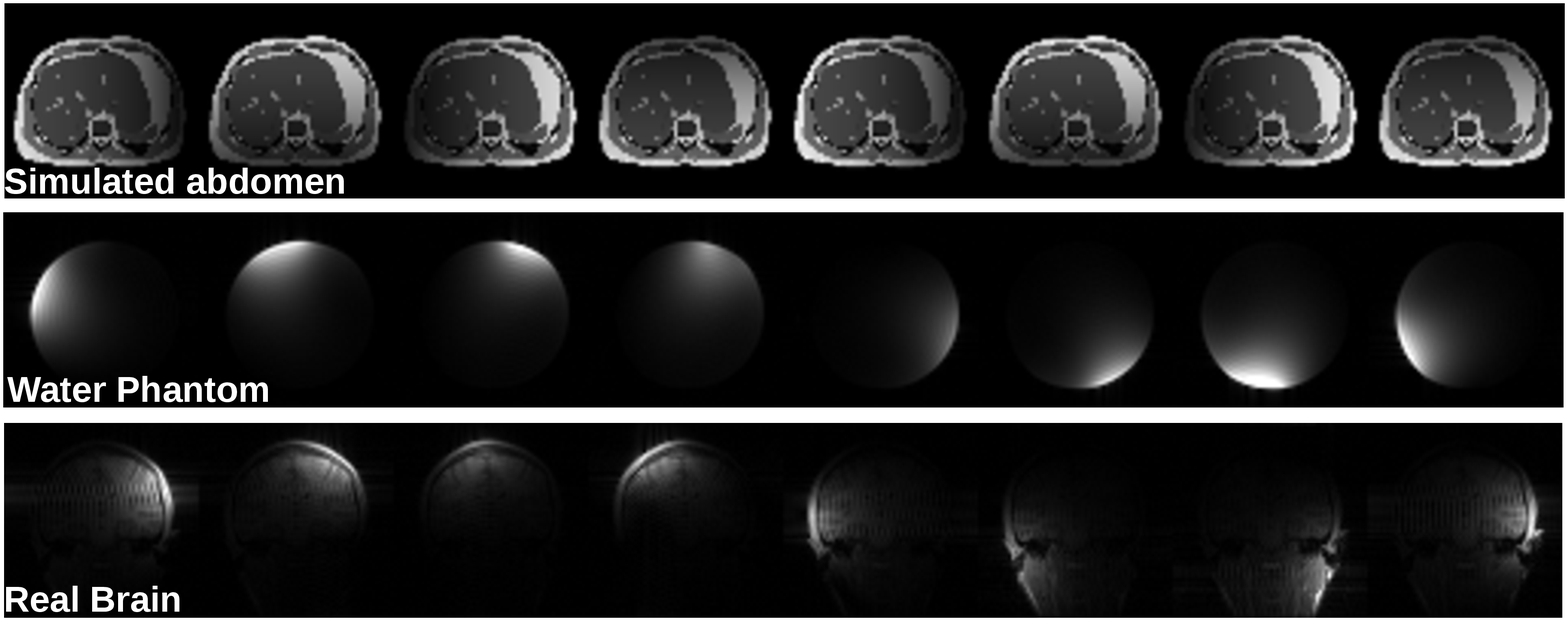}
\caption{The first three rows show the datasets (coil-by-coil images) used for the Monte--Carlo simulation, the phantom and the \emph{in--vivo} experiments. Due to space constraints, we only show 8 of the 32 channels of the central slice for each dataset.}
\label{fig:datasets}
\end{figure}

From each of the first two data sets, four sub-sampling scenarios were considered (see Fig.\ref{fig:patterns}).

\begin{enumerate}
	\item First, we considered a CAIPIRINHA type sub-sampling pattern along the two phase--encoding dimensions with an acceleration rate R=2. A fully--sampled rectangular ACS region in the center of the \bk--space was acquired containing $8\times 4$ fully sampled lines along the frequency--encoding direction (CAIPI-Rect). We used a GRAPPA kernel of size  $\left[ 3,3,3 \right]$ to reconstruct the missing data. 
	\item Second, we studied the case of an ellipsoidal ACS region over the same CAIPIRINHA type sub-sampling pattern (CAIPI-Ellip). The axis of the ellipsoidal ACS were of size $8\times 4$ and the reconstruction kernel size was kept at $\left[ 3,3,3 \right]$. 
	\item Third, we considered the case of using a variable density sampling as in \cite{Heidemann07}. We divided the \bk--space into three circular regions. The center was a fully--sampled ellipsoid (ACS region) with axis $8\times 4$. Then, a middle region with axis $6\times 320$ was sub-sampled at a rate $R=2$ and finally the outer region covering the corners was sub-sampled at a rate $R=3$. The inner region was reconstructed with a $\left[ 3,3,3 \right]$ kernel, whereas for the outer region we used a $\left[ 5,5,3 \right]$ kernel (CAIPI-VD).
	\item Fourth, we considered the case of a random sub-sampling pattern along the two phase--encoding directions with acceleration rate $R=2$, preserving a low-frequency ACS region of size $8\times 4$. Again, a $\left[ 3,3,3 \right]$ kernel was used for reconstruction. The method will be hereafter referred to as Random. 
\end{enumerate}

\begin{figure}[tb]
\centering
\includegraphics[width=0.99\columnwidth]{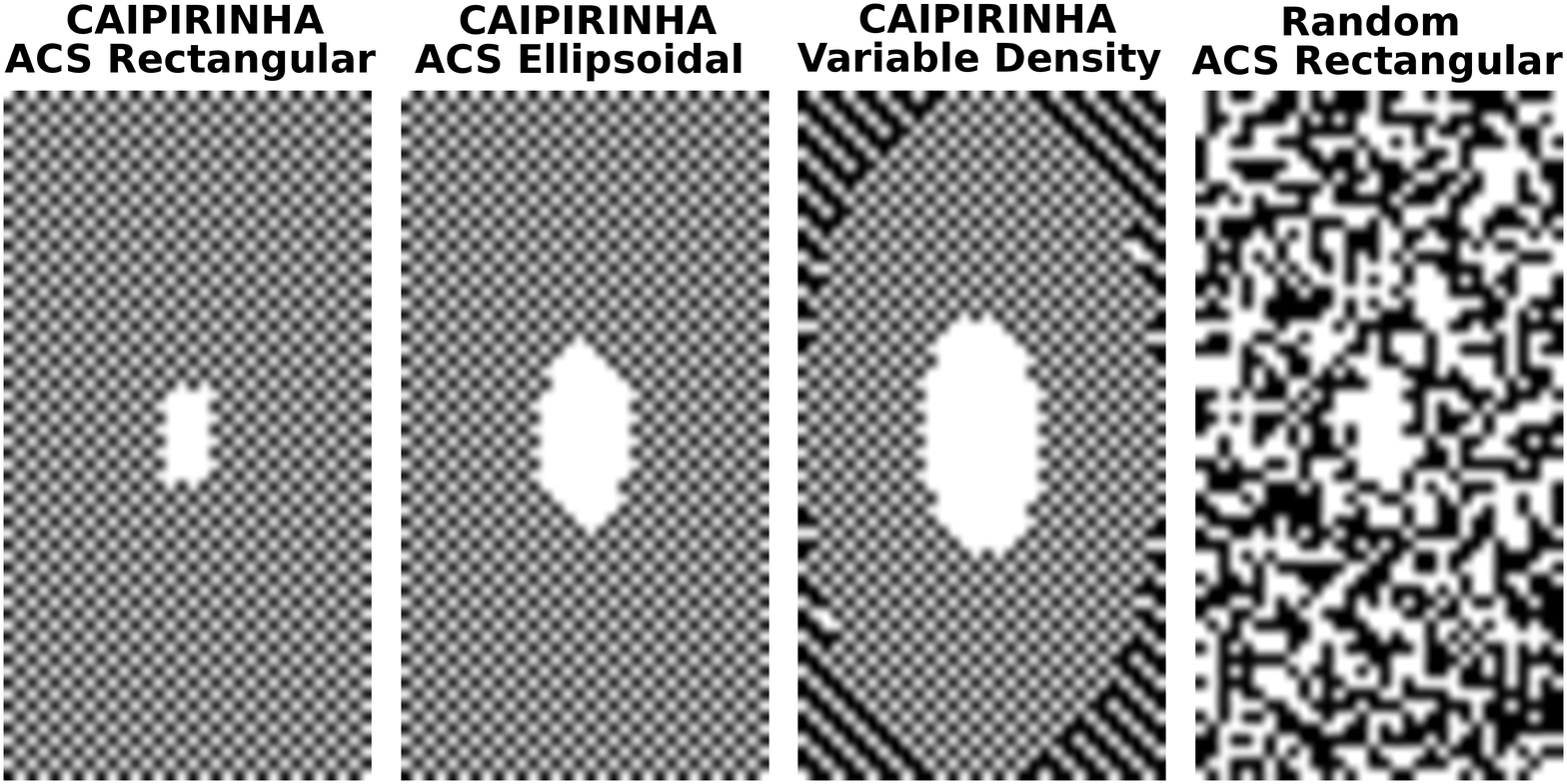}
\caption{Sub-sampling patterns used for the phantom experiments, all containing a central fully--sampled ACS region.}
\label{fig:patterns}
\end{figure}

For these experiments, $g$-factor maps were obtained in two different ways. First, we followed a Monte--Carlo approach to calculate the sample standard deviation for each pixel from all the realizations for both the unaccelerated and the accelerated images. The $g$-factor maps are computed using the definition in Eq.~\eqref{eq:$g$-factor_def}. Second, we directly computed the $g$-factor maps using the proposed \bk--space method as in Eq.~\eqref{eq:$g$-factor_final}. Our method requires as inputs the \bk--space noise matrices ${\bf \Gamma}$ and ${\bf C}$ in the sub-sampled images, the GRAPPA kernel and the coil-combination vector. For the simulated scenario, these parameters were known. For the acquired phantom, these matrices needed to be estimated. Using the multiple realizations of the same acquisition that were available we simply estimated them as the sample covariance matrices obtained across realizations. Furthermore we exploited the stationarity accross the image by computing them for every location and averaging afterwards. 

Finally, in order to prove the applicability to clinical scenarios, we studied the impact of the sub-sampling pattern using the \emph{in--vivo} scan. Three different sub-sampling patterns were considered at an acceleration rate of $R=4$: Rectangular type ($R=2\times 2$), CAIPIRINHA-type ($R=2\times 2$) and Uniformly Random, as shown in Fig.\ref{fig:InVivo}. The acquired \bk--space was sub-sampled preserving an $8\times 4$ fully sampled ACS region. In this case we needed to estimate the noise from a single repetition, which we did using the Mean Absolute Deviation estimator after a wavelet filtering the preserves the high-frequency band, which is supposed to contain only the noise, as in \cite{AjaNoiseBook}.

All image reconstruction and $g$-factor maps estimation were performed using Matlab and run on a shared computer with an Intel\textregistered Xeon\textregistered CPU E5-2695 v3 @2.30GHz
Processor and 110 GB of RAM. In the spirit of reproducible research, we provide a software package including both the data sets and the code that we used, allowing to reproduce all the results included in this manuscript. This package can be downloaded from \url{http://lpi.tel.uva.es/grappa_kspace}. 

%%%%%%%%%%%%%%%%%%%%%%%%%%%%%%%%%%%%%%%%%%%%%%%%%%%%%%%%
%RESULTS---------
%%%%%%%%%%%%%%%%%%%%%%%%%%%%%%%%%%%%%%%%%%%%%%%%%%%%%%%%

\section{Results}

Fig.~\ref{fig:SyntheticResults} shows the $g$-factor for each pixel at the intermediate slice of the 3D simulated dataset reconstructed using 3D-GRAPPA with the four sub-sampling patterns described above. The first two columns show the Monte--Carlo empirical $g$-factor maps and the analytical \bk--space $g$-factor maps obtained with our method, respectively. The third colum shows the differences between the analytical estimation against the Monte Carlo reference. Finally, the last column shows a boxplot diagram of absolute error distribution along the third dimension (along the slices). Fig.~\ref{fig:RealResults} shows the same results obtained for the water phantom. 

\begin{figure}[tb]
\centering
\includegraphics[width=1\columnwidth]{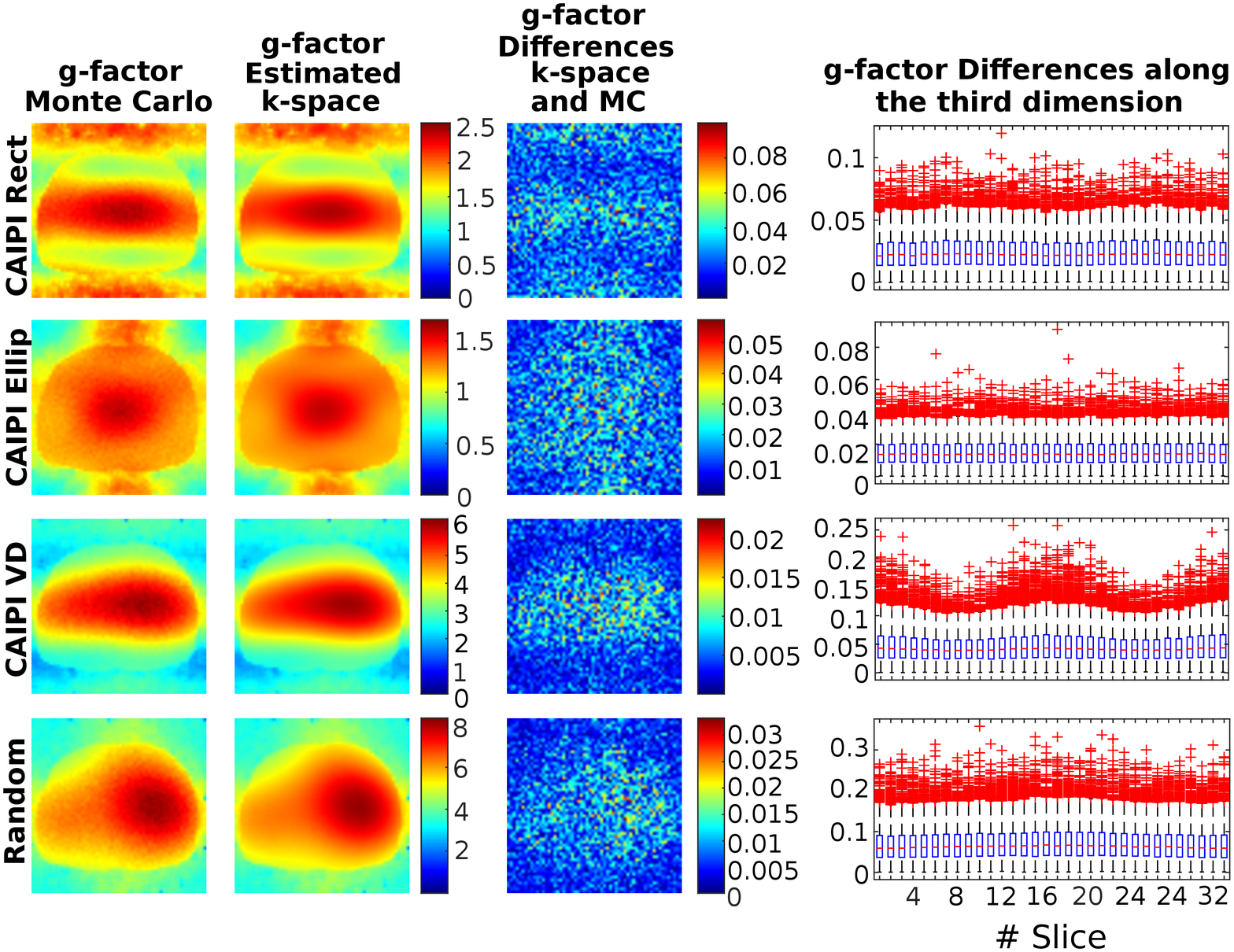}
\caption{$g$-factor maps for the synthetic phantom obtained through a Monte--Carlo strategy (first column) and estimated using the proposed \bk--space method (second column) for the intermediate slice of the 3D volume. The third column shows the absolute differences between the \bk--space method against the Monte Carlo estimation. The last column shows the absolute error distribution along the third dimension. Several scenarios are studied: CAIPIRINHA--type with rectangular ACS region (first row), CAIPIRINHA--type with ellipsoidal ACS region (second row), CAIPIRINHA--type with Variable Density sampling (third row) and random sub-sampling with rectangular ACS region (fourth row).}
\label{fig:SyntheticResults}
\end{figure}

\begin{figure}[tb]
\centering
\includegraphics[width=1\columnwidth]{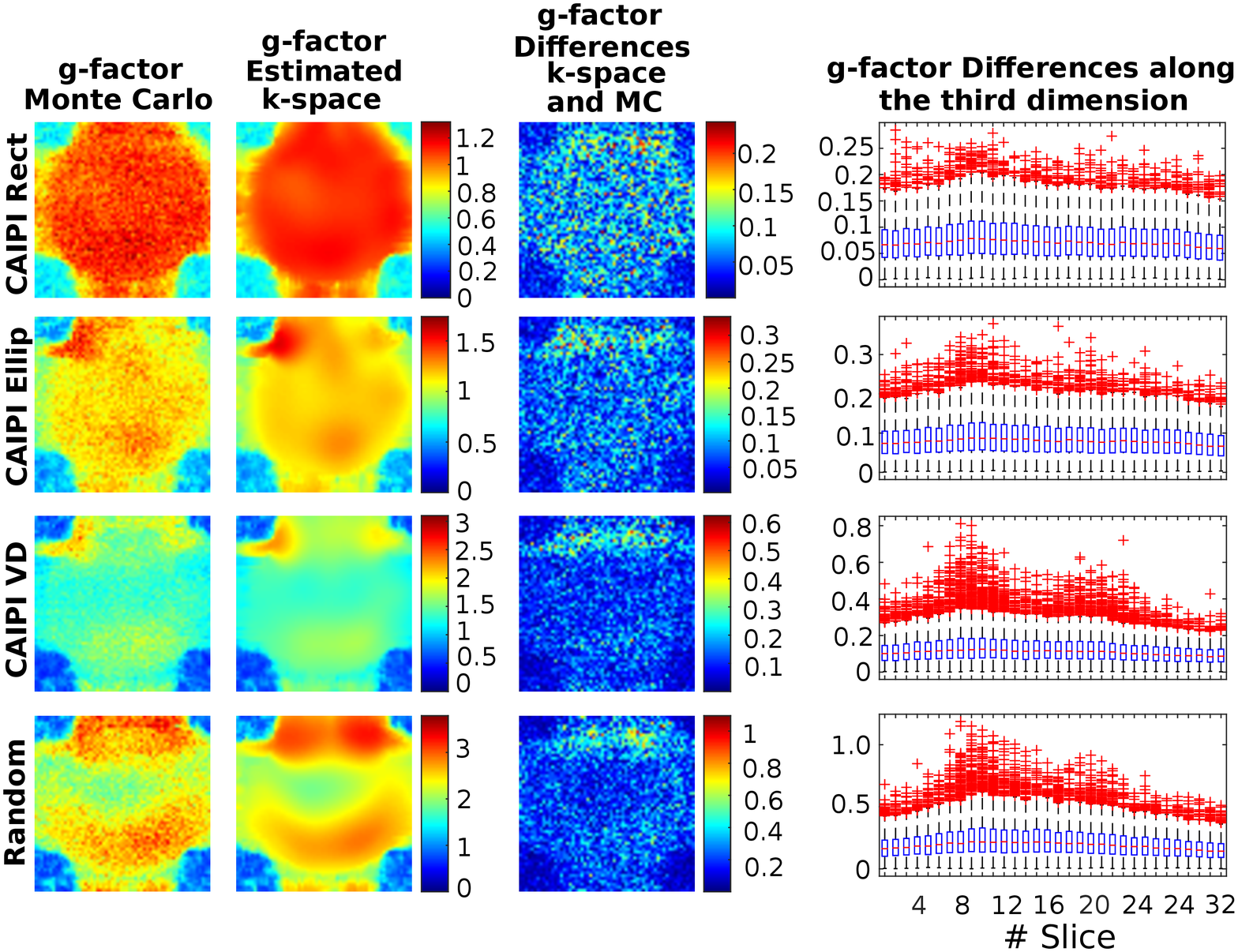}
\caption{$g$-factor maps for the scanned water phantom obtained through a Monte--Carlo strategy (first column) and estimated using the proposed \bk--space method (second column). The third column shows the absolute differences between the \bk--space method against the Monte Carlo estimation. The last column shows the absolute error distribution along the third dimension. Same scenarios as for the synthetic phantom are shown.}
\label{fig:RealResults}
\end{figure}

As can be seen, the analytical method is able to provide an accurate estimation of the $g$-factor maps under all the sub-sampling scenarios tested, both for the synthetic and real phantom. The differences observed between the $g$-factor maps corresponding to the analytical approach and the Monte--Carlo approach in the synthetic experiments can be atributed to the approximate behaviour inherent to Monte--Carlo methods, since they need an infinite number of realizations to be exact. Different realizations of the entire experiment did not show any systematic deviations between our solution and that from Monte--Carlo.

Higher differences between both maps can be observed for the real experiments, which is consistent with the reduced number of realizations available for the real scenario (100 realizations) compared to the simulated one (4000 realizations). Furthermore, a slight variability can be observed in the differences over the slices. 

In order to show that the \bk--space analysis can be applied to \emph{in--vivo} datasets, the $g$-factor maps obtained with different sub-sampling patterns are shown in Fig.\ref{fig:InVivo}. In the studied case, in terms of noise the choice between a Rectangular--type pattern and a CAIPIRINHA--type pattern does not seem to have a significant impact. However, we observe that using a uniform random pattern results in a higher noise amplification factor.

%% DHA: I would suggest making the font size substantially bigger for all labels in this image, even if you need two lines per label. 

\begin{figure}[tb]
\centering
\includegraphics[width=0.5\textwidth]{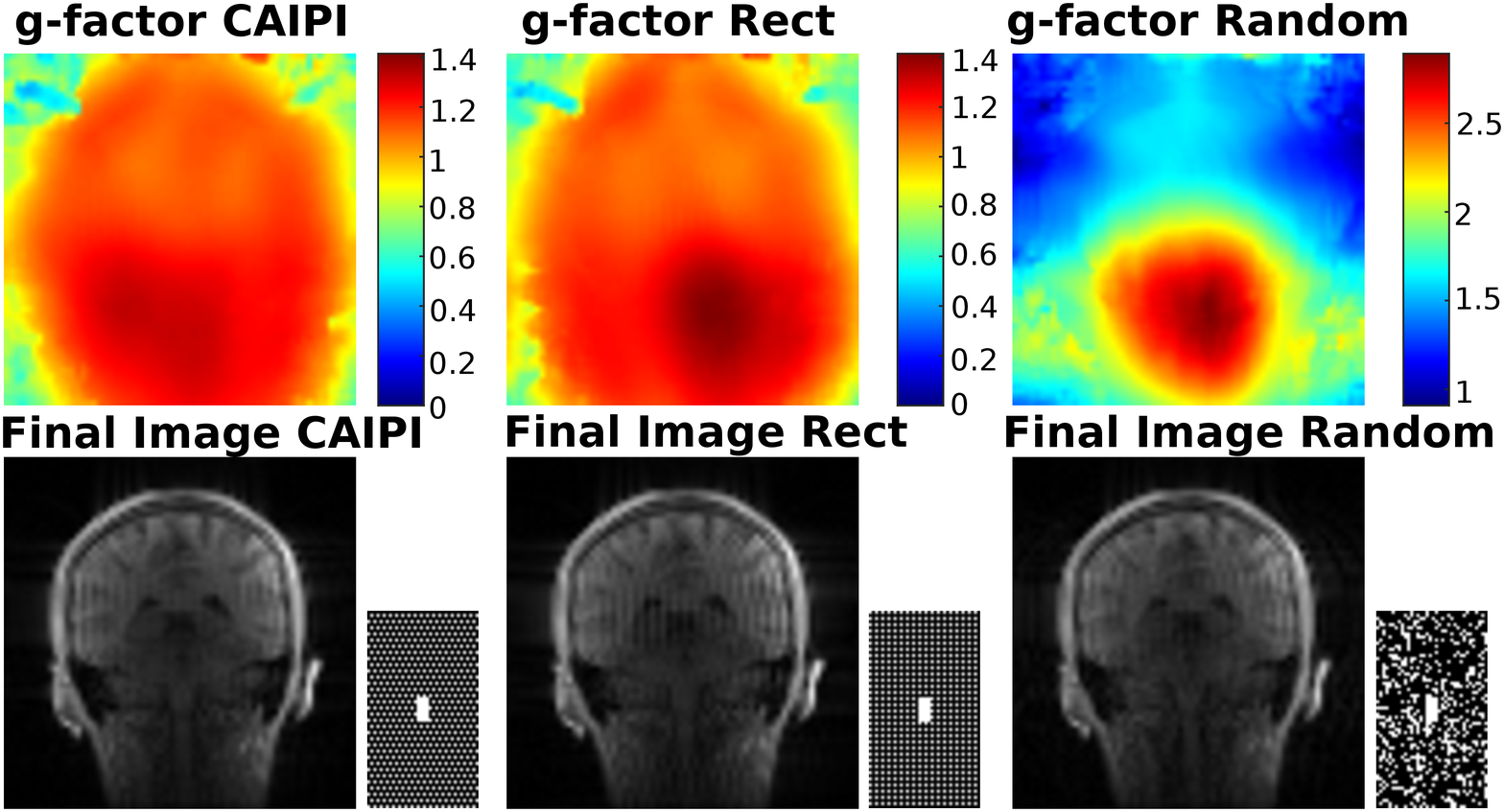}
\caption{Calculated $g$-factor maps for the \emph{in--vivo} brain experiment obtained through the proposed \bk--space method. Three different subsampling patterns are studied at the same acceleration rate $R=4$: Rectangular type ($R=2\times 2$), CAIPIRINHA-type ($R=2\times 2$) and Uniformly Random. $g$-Factor maps (first row) and final composite coil-combined images (second row) for the intermediate slice of the 3D volume. Next to each coil-combined image we show the subsampling pattern used along the two phase--encoding dimensions.}
\label{fig:InVivo}
\end{figure}

Finally, as for computational load, we include the computation time to obtain the $g$-factor maps for each of the sub-sampling patterns previously described in Table~\ref{table:times}. As expected, higher numbers of columns with unique patterns result in higher computational time requirements. 

\begin{table}[h]
	\centering
	\caption{Computation times (seconds)}\label{table:times}
    \begin{tabular}{| c | c | c | c |}
    \hline
    CAIPI Rect & CAIPI Ellip & CAIPI VD & Random \\ \hline
    121.32 & 215.28 & 688.71 & 629.08 \\ \hline
    \hline
    \end{tabular}
\end{table}

%%%%%%%%%%%%%%%%%%%%%%%%%%%%%%%%%%%%%%%%%%%%%%
%DISCUSSION------------------------------------------
%%%%%%%%%%%%%%%%%%%%%%%%%%%%%%%%%%%%%%%%%%

\section{Discussion}

In this work, we have proposed an analytical method for noise characterization in 3D-GRAPPA reconstructions by extending  a previously proposed 2D method~\cite{Rabanillo17}. Our method provides an exact characterization of noise under two common assumptions: stationarity and uncorrelation of noise in the acquired \bk--space. This is a challenging problem due to the very large size of the covariance matrices involved in the noise propagation. To overcome this computational burden and still provide an exact characterization, our analysis relies on two cornerstones. First, instead of operating in the image--space based on approximately--equivalent reconstructions, it directly operates in the \bk--space in order to account for all the \bk--space correlations by characterizing the noise properties along the entire reconstruction pipeline. Second, our analysis exploits the extensive symmetries and the separability in the reconstruction steps, which allows us to operate with much smaller matrices. As a result, the proposed method provided an accurate characterization of noise in 3D sub-sampled acquisitions reconstructed with GRAPPA under the assumptions of stationarity and uncorrelation in the original \bk--space sub-sampled acquisition.

We have illustrated the performance of the proposed method under various sub-sampling patterns (Rectangular, CAIPIRINHA, Random and Variable-Density) and different ACS regions. Importantly, 3D acquisitions enable a high degree of flexibility in sub-sampling the \bk--space, therefore many other sub-sampling patterns could benefit from our method. Furthermore, our method may be useful in optimizing the GRAPPA reconstruction kernel size in terms of noise performance, since it has recently been proven that the kernel choice can have a direct impact both on the quality of the image and on the extraction of quantitative measures~\cite{GRAPPAfMRI}.

This study presents some limitations. First, the analysis focuses on GRAPPA reconstruction followed by a linear coil combination such as \cite{Walsh00}. If SoS is used instead, the noise distribution can be approximated as a non-Central $\chi$ distribution~\cite{AjaMRM11}, whose effective parameters need to be calculated. However, noise characterization of SoS coil combination is straightforward from the final covariance matrices ${\bf \Gamma}_6(x)$ and ${\bf C}_6(x)$ by simply applying them to eq.(17-18) from \cite{AjaMRM11}. Second, as in Ref.~\cite{Rabanillo17}, we have assumed that the kernel used for reconstruction is independent of  noise in the acquired image, which is not strictly true when the kernel is autocalibrated from the data instead of pre-calibrated from a separate pre-scan. However, this assumption is expected to be a good approximation due to the typical overdetermination of GRAPPA weights estimation from the ACS region, and for this reason it has become a common assumption in the literature. Third, our analysis is focused on providing the final voxel--wise $g$-factor maps. However, for certain applications such as optimal filtering \cite{Sprenger16} it may be useful to additionally characterize  the noise correlation between different locations in the reconstructed image. Importantly, our proposed method can be extended to provide the correlation between any two points in the final image, thus allowing to obtain cross-correlation maps.

Although this manuscript does not directly consider non-Cartesian 3D acquisitions, the proposed techniques can also be extended to non-Cartesian acquisitions (eg: radial or spiral trajectories) if preceded by a linear interpolation into a Cartesian grid. Such an interpolation would introduce  correlations in the sub-sampled \bk--space and may introduce non-stationarity if the interpolation weights are not uniform across the \bk--space. In such scenarios, it would suffice to take into account this non-stationarity when building the correlation matrices of the sub-sampled Cartesian \bk--space at the first step of our analysis. However, the increased correlations in the \bk--space may result in longer computation times. A potential scenario where our method could be applied is  non-Cartesian acquisitions regridded with the GROG operator, as in \cite{Seiberlich08}, since no extra-correlations would be introduced and only the non-stationarity should be taken into account. Finally, the Matlab code used in this work has not been optimized for speed. Indeed, significant acceleration may be achieved with an implementation that exploits the high degree of paralelism present in several reconstruction steps within our method. 

\section{Conclusion}

We have extended the \bk--space noise analysis proposed in \cite{Rabanillo17} to characterize noise in 3D-GRAPPA reconstructions of volumes sub-sampled along two phase--encoding directions. By directly operating in \bk--space our analysis is able to exactly characterize the noise under the assumptions of stationarity and uncorrelation in the acquired \bk--space. By exploiting both the symmetry and separability in the reconstruction steps, the proposed method enables efficient and exact noise characterization in 3D-GRAPPA . 

\appendix
\section{Proof that shifted columns are equivalent} 
\label{appendix}

In this appendix we will present the proof that two columns that present the same pattern except for a shift provide a matrix ${\bf \Gamma}_3^{m_{uq}}$ with the same diagonal. We can write the correlation matrix ${\bf \Gamma}$ of a reference column with another column as follows:
\begin{equation} \label{eq:GammaRef}
{\bf \Gamma}_{a,1} = \left( \begin{array}{cccc}
\vec{{\bf c}_{1}}[n] & \vec{{\bf c}_{2}}[n] & \cdots & \vec{{\bf c}_{N_x}}[n]\\
\end{array}\right),
\end{equation}
where the vectors $\vec{{\bf c}_{i}}[n], n=\{0\ldots N_x-1\}$ denote the columns of this matrix. Let us consider a different reference column that correlates with a surrounding one as a shifted version of the already described reference column. We will initially consider a shift of a single element. Consequently, its correlation matrix can be described as:
\begin{equation} \label{eq:GammaShift}
{\bf \Gamma}_{b,1} = \left( \begin{array}{cccc}
\vec{{\bf c}_{N_x}}[n-1] & \vec{{\bf c}_{1}}[n-1] & \cdots & \vec{{\bf c}_{N_x-1}}[n-1]\\
\end{array}\right).
\end{equation}
It should be noted that $\vec{{\bf c}_{i}}[n-1], n=\{0\ldots N_x-1\}$ denotes a circular shift of the column vector. Computing the iFFT along the column dimension results in operating as follows on the ${\bf \Gamma}$ matrix, as stated in Eq.~\eqref{eq:GammaHybridCol}:
\begin{equation} \label{eq:BasicFFT}
{\bf \Gamma}_2 = {\bf F}_I \cdot {\bf \Gamma}_1 \cdot {\bf F}_I^H=[{\bf F}_I \cdot( {\bf F}_I  \cdot {\bf \Gamma}_1)^H]^H =[{\bf F}_I \cdot {\bf \Psi}^H]^H .
\end{equation}
where we have defined an intermediate matrix ${\bf \Psi}={\bf F}_I  \cdot {\bf \Gamma}_1$. Defining this intermediate matrix for the first reference column as:
\begin{equation} \label{eq:GammaRef}
{\bf \Psi}_{a} = \left( \begin{array}{cccc}
\vec{{\bf d}_{1}}[n] & \vec{{\bf d}_{2}}[n] & \cdots & \vec{{\bf d}_{N_x}}[n]\\
\end{array}\right),
\end{equation}
and using the Shift Theorem of the Fourier Transform, it is inmediate to observe that:
\begin{equation} \label{eq:GammaRef}
{\bf \Psi}_{b} = \left( \begin{array}{cccc}
\vec{{\bf d}_{N_x}}[n] \cdot\vec{{\bf w}}[n]  & \vec{{\bf d}_{1}}[n]\cdot\vec{{\bf w}}[n]  & \cdots & \vec{{\bf d}_{N_x-1}}[n]\cdot\vec{{\bf w}}[n] \\
\end{array}\right),
\end{equation}
where $\vec{{\bf w}}[n]=exp\left(\iota \frac{2\cdot\pi\cdot n}{N_x}\right), n=\{0\ldots N_x-1\}$. To avoid confusion with the indices, we have denoted the imaginary unit by $\iota$. If we denote the rows of the matrix ${\bf \Psi}_{a}$ as $\vec{{\bf r}_{j}^a}[n], n=\{0\ldots N_x-1\}$, we can see that the rows of the matrix ${\bf \Psi}_{b}$ can be expressed as $\vec{{\bf r}_{j}^b}[n]=\vec{{\bf r}_{j}^a}[n-1]\cdot exp\left(\iota\frac{2\cdot\pi\cdot j}{N_x}\right)$. When we compute the final matrix ${\bf \Gamma}_2$, we operate on ${\bf \Psi}^H$, that is to say, on the rows of ${\bf \Psi}$. Using again the Shift Theorem, we can see that:
\begin{equation} \label{eq:BasicFFT}
{\bf \Gamma}_{b,2} = {\bf \Gamma}_{b,1} \circ exp\left(\iota \frac{2\cdot\pi\cdot (i-j)}{N_x}\right), 
\end{equation}
where $\{i,j\}=\{0\ldots N_x-1\}$ denote the row and column of the matrix. After operating along the first dimension, we only care about the correlations of a reference point with other points that are in the same plane (equal position in the first dimension). This means that we only need to keep the diagonal of the previous matrix, as already explained. Comparing the diagonals of these two matrices, we can clearly see that $\text{diag}({\bf \Gamma}_{b,2})=\text{diag}({\bf \Gamma}_{b,2})$. Finally, we have derived this proof for the case of shift of one element. For the general case of a shift of any size, since it can be seen as the successive application of many one-element shifts, the previous equality will hold.

\section*{Acknowledgments}
The authors acknowledge MICIN for grants TEC2013-44194P, TEC 2014-57428 and TEC2017-82408-R, as well as Junta de Castilla y León for grant VA069U16. The first author acknowledges MINECO for FPI grant BES-2014-069524.

\bibliographystyle{mrm}
\bibliography{strings,grappa}

\end{document}